\documentclass[onecolumn,preprint,pra]{revtex4}
\usepackage{epsf}
\usepackage{amsmath,amsthm,amssymb}
\usepackage{color}
\usepackage{dcolumn}
\usepackage{bm}
\usepackage{graphicx,epsfig}
\usepackage[dvipsnames]{xcolor}

\graphicspath{{fig/}}
\everymath{\displaystyle}

\begin{document} 

\title{Twisted particles in heavy-ion collisions}

\author{Alexander J. Silenko$^{1,2,3}$}
\email{alsilenko@mail.ru}
\author{Pengming Zhang$^{4}$}
\email{zhangpm5@mail.sysu.edu.cn}
\author{Liping Zou$^{5,2}$}
\email{zoulp@impcas.ac.cn}

\affiliation{$^1$Bogoliubov Laboratory of Theoretical Physics, Joint
Institute for Nuclear Research,
Dubna 141980, Russia}
\affiliation{$^2$Institute of Modern Physics, Chinese Academy of
Sciences, Lanzhou 730000, China}
\affiliation{$^3$Research Institute for Nuclear Problems, Belarusian
State University, Minsk 220030, Belarus}
\affiliation{$^4$School of Physics and Astronomy, Sun Yat-sen University,
519082 Zhuhai, China}
\affiliation{$^5$Sino-French Institute of Nuclear Engineering and Technology, Sun Yat-Sen University,
519082 Zhuhai, China}



\begin{abstract}
The importance of production of twisted (vortex) particles in heavy-ion collisions is analyzed. Free twisted particles can possess giant intrinsic orbital angular momenta. Twisted particles are spatially localized and can be rather ubiquitous in laboratories and nature. Twisted photons have nonzero effective masses. Charged twisted particles can be recognized by their dynamics, magnetic moments, and specific effects in external fields.
\end{abstract}
\maketitle

Twisted (vortex) particles possess an intrinsic orbital angular momentum (OAM) and can be presented by wave beams or packets. Wave beams and packets are localized with respect to two and three dimensions and are described by two and three discrete transverse quantum numbers, respectively. Free twisted particle beams of photons, electrons, and neutrons can be described in the cylindrical coordinates by the Laguerre-Gauss wave function
\begin{equation}
\begin{array}{c}
\Psi={\cal A}\exp{(i\Phi)},\qquad
{\cal A}=\frac{C_{n\ell}}{w(z)}\left(\frac{\sqrt2r}{w(z)}\right)^{|\ell|}
L_n^{|\ell|}\left(\frac{2r^2}{w^2(z)}\right)\exp{\left(-\frac{r^2}{w^2(z)}\right)},\\
\Phi=\ell\phi+\frac{kr^2}{2R(z)}-(2n+|\ell|+1)\varphi(z),\quad
C_{n\ell}=\sqrt{\frac{2n!}{\pi(n+|\ell|)!}},\quad w(z)=w_0\sqrt{1+\frac{4z^2}{k^2w_0^4}},\\
R(z)=z+\frac{k^2w_0^4}{4z},\qquad
\varphi(z)=\arctan{\left(\frac{2z}{kw_0^2}\right)},\qquad
\int{\Psi^\dag\Psi r\,dr\,d\phi}=1,
\end{array}\label{eqint}
\end{equation}
where the real functions ${\cal A}$ and $\Phi$ define the amplitude and phase of the beam,
$k$ is the beam wavenumber, $w_0$ is the minimum beam
width, $L_n^{|\ell|}$ is the generalized Laguerre polynomial, $\hbar\ell$ is the OAM, and
$n = 0, 1, 2,
\dots$ is the radial quantum number. Free twisted particles can possess giant intrinsic OAMs. Photons with OAMs more than 10000$\hbar$ \cite{PNAS} and electrons with OAMs up to 1000$\hbar$ \cite{VGRILLO} have been obtained. 
Twisted photons have nonzero \emph{effective} masses \cite{photonPRA}. Charged twisted particles can be recognized by their dynamics \cite{Manipulating,ResonanceTwistedElectrons}, magnetic moments \cite{BliokhSOI}, and specific effects in external fields \cite{PhysRevLettLanzhou2019,snakelike,arXiv}.

The importance of production of twisted particles at heavy-ion collisions (HIC) is evident. The similar effect of the global polarization of produced particles \cite{GlPolar} attracts a lot of attention. To study the above-mentioned problem at noncentral HIC, one should take into account an appearance of a strong magnetic field \cite{SkokovIllarionovToneev} and a fast rotation of nuclear matter leading to its large vorticity \cite{Baznatetal}. 

The recent theoretical \cite{theoretic} and experimental \cite{experimental} papers unambiguously show that the photons radiated by electrons in a helical or circular motion are twisted (i.e., have nonzero OAMs). The fast rotation of nuclear matter at noncentral HIC leads to a charge rotation and, therefore, to an emission of twisted photons. We suppose that other particles can also be produced in twisted states. The vorticity of nuclear matter is, to some extent, an effect which is similar to the twist and leads to a production of particles in twisted states \cite{footnote}. To describe processes taking place in a rotating nuclear matter, one often uses uniformly rotating frames (see, e.g., Ref. \cite{Baznatetal}). For a quantum-mechanical description of particles in such frames, it is convenient to apply the Foldy-Wouthuysen (FW) representation providing for the Schr\"{o}dinger picture of relativistic quantum mechanics (see Refs. \cite{PRA2015,PRAFW} and references therein). The relativistic FW Hamiltonian describing a spin-1/2 particle in the frame uniformly rotating with the angular velocity $\bm \omega$ has the form \cite{PRD2}
\begin{equation}
\begin{array}{c}
{\cal H}_{FW}=\beta\sqrt{m^2+\bm p^2}-\bm \omega\cdot(\bm L+\bm s),
\end{array}\label{eqrotfr}
\end{equation}
where $\beta$ is the Dirac matrix and $\bm L$ and $\bm s$ are the OAM and spin operators. The eigenstates of this operator have definite integer OAMs and the corresponding eigenfunctions describe Laguerre-Gauss beams.

The above-mentioned arguments allow us to predict that the production of twisted photons and twisted massive particles with different spins should be usual in noncentral heavy-ion collisions. The production of charged twisted particles can be stimulated, besides the fast rotation of the nuclear matter, by the strong magnetic field. Our analysis based on precedent studies \cite{Baznatetal,theoretic,experimental,PRD2} unambiguously shows that the production of twisted particles in noncentral heavy-ion collisions is rather widespread.

\textbf{\emph{Acknowledgments}.} The authors are grateful to O.V. Teryaev for multiple discussions of the problem and valuable comments.
The work was supported by the National Natural Science
Foundation of China (grant No. 11975320 and No. 11805242) and by the Chinese Academy of Sciences President's International Fellowship Initiative (grant No. 2019VMA0019).
A. J. S. also acknowledges hospitality and support by the
Institute of Modern
Physics of the Chinese Academy of Sciences. 


\begin{thebibliography}{}

\bibitem{PNAS} R. Fickler, G. Campbell, B. Buchler, Ping Koy Lam, and A. Zeilinger, Quantum entanglement of angular momentum states with quantum numbers up to 10,010,
Proc. Natl. Acad. Sci. U.S.A. \textbf{113}, 13642 (2016). 

\bibitem{VGRILLO} E. Mafakheri \emph{et al.}, 
Realization of electron vortices with large orbital angular momentum using miniature holograms fabricated by electron beam lithography, 
Appl. Phys. Lett. \textbf{110}, 093113 (2017). 

\bibitem{photonPRA}
A. J. Silenko, Pengming Zhang, and Liping Zou, 
Relativistic quantum-mechanical description of twisted paraxial electron and photon beams, 
Phys. Rev. A \textbf{100}, 030101(R) (2019). 

\bibitem{Manipulating}
A. J. Silenko, Pengming Zhang and Liping Zou, Manipulating Twisted Electron Beams, 
Phys. Rev. Lett. \textbf{119}, 243903 (2017). 

\bibitem{ResonanceTwistedElectrons}
A. J. Silenko, Pengming Zhang and Liping Zou, Relativistic quantum dynamics of twisted electron beams in arbitrary electric and magnetic fields, 
Phys. Rev. Lett. \textbf{121}, 043202 (2018). 

\bibitem{BliokhSOI}
K. Y. Bliokh  \emph{et al.}, 
Theory and applications of free-electron vortex states, 
Phys. Rep. \textbf{690}, 1 (2017). 
 
\bibitem{PhysRevLettLanzhou2019}
A. J. Silenko, Pengming Zhang, and Liping Zou, Electric Quadrupole Moment and the Tensor Magnetic Polarizability
of Twisted Electrons and a Potential for their Measurements, 
Phys. Rev. Lett. \textbf{122}, 063201 (2019). 

\bibitem{snakelike}
A. J. Silenko and O. V. Teryaev, Siberian Snake-Like Behavior for an Orbital Polarization of a Beam of Twisted (Vortex) Electrons,
Phys. Part. Nucl. Lett. \textbf{16}, 77 (2019). 

\bibitem{arXiv}
Liping Zou, Pengming Zhang, and A. J. Silenko, General quantum-mechanical solution for twisted electrons in a uniform magnetic field, \textbf{103}, L010201 (2021). 

\bibitem{GlPolar}
Z.-T. Liang, M. A. Lisa, and X.-N. Wang, Global Polarization Effect in the Extremely Rapidly Rotating QGP in HIC,
Nucl. Phys. News, \textbf{30}, 10 (2020); 
L. Adamczyk et al., Global $\Lambda$ hyperon polarization in nuclear collisions, 
Nature \textbf{548}, 62 (2017); 
Q.  Wang, Global and local spin polarization in heavy ion collisions: a brief overview, Nucl. Phys. A \textbf{967}, 225 (2017). 

\bibitem{SkokovIllarionovToneev}
V. V. Skokov, A. Yu. Illarionov and V. D. Toneev, Estimate of the magnetic field strength in heavy-ion collisions, 
Int. J. Mod. Phys. A \textbf{24}, 5925 (2009). 

\bibitem{Baznatetal}
M. Baznat, K. Gudima, A. Sorin, and O. Teryaev, Helicity separation in heavy-ion collisions, 
Phys. Rev. C \textbf{88}, 061901(R) (2013); Femto-vortex sheets and hyperon polarization in heavy-ion collisions, Phys. Rev. C \textbf{93}, 031902(R) (2016); Hyperon polarization in heavy-ion collisions and holographic gravitational anomaly, Phys. Rev. C \textbf{97}, 041902(R) (2018); A. Zinchenko, A. Sorin, O. Teryaev, and M. Baznat, Vorticity structure and polarization of $\Lambda$ hyperons in heavy-ion collisions, J. Phys.: Conf. Ser. \textbf{1435}, 012030 (2020). 

\bibitem{theoretic}
M. Katoh \emph{et al.}, Angular Momentum of Twisted Radiation from an Electron in Spiral Motion, 
Phys. Rev. Lett. \textbf{118}, 094801 (2017); S. V. Abdrashitov, O. V. Bogdanov, P. O. Kazinski, and T. A. Tukhfatullin, 
Orbital angular momentum of channeling radiation from relativistic electrons in thin Si crystal, 
Phys. Lett. A \textbf{382}, 3141 (2018); V. Epp, J. Janz, and M. Zotova, Angular momentum of radiation at axial channeling, 
Nucl. Instrum. Methods Phys. Res., Sect. B \textbf{436}, 78 (2018); M. Katoh \emph{et al.}, Helical Phase Structure of Radiation from an Electron in Circular Motion, 
Sci. Rep. \textbf{7}, 6130 (2017); O. V. Bogdanov, P. O. Kazinski, and G. Yu. Lazarenko, Probability of radiation of twisted photons by classical currents, 
Phys. Rev. A \textbf{97}, 033837 (2018); Semiclassical probability of radiation of twisted photons in the ultrarelativistic limit, Phys. Rev. D \textbf{99}, 116016 (2019); Probability of radiation of twisted photons in the infrared domain, Ann. Phys. (N.Y.) \textbf{406}, 114 (2019); V. Epp, U. Guselnikova, Angular momentum of radiation from a charge in circular and spiral motion, 
Phys. Lett. A \textbf{383}, 2668 (2019). 

\bibitem{experimental}
M. Katoh \emph{et al.}, Helical Phase Structure of Radiation from an Electron in Circular Motion, 
Sci. Rep. \textbf{7}, 6130 (2017); T. Kaneyasu \emph{et al.}, Observation of an optical vortex beam from a helical undulator in the XUV region,
J. Synchrotron Rad. \textbf{24}, 934 (2019).

\bibitem{footnote}
The authors are indebted to O.V. Teryaev for clarification of this point and suggestion to consider twisted states in heavy-ion collisions as a counterpart of classical vorticity.

\bibitem{PRA2015}
A.\,J. Silenko, 
General method of the relativistic Foldy-Wouthuysen transformation and proof of validity of the Foldy-Wouthuysen Hamiltonian,
Phys. Rev. A \textbf{91}, 022103 (2015). 

\bibitem{PRAFW}
Liping Zou, Pengming Zhang, and A. J. Silenko, Position and spin in relativistic quantum mechanics, 
Phys. Rev. A \textbf{101}, 032117 (2020). 

\bibitem{PRD2}
A. J. Silenko and O. V. Teryaev, 
Equivalence principle and experimental tests of gravitational spin effects,
Phys. Rev. D {\bf 76}, 061101(R) (2007).

\end{thebibliography}
\end{document}